# A COMPACT HIGH-ENERGY NEUTRON SPECTROMETER

F.D. Brooks[1,*], A. Buffler[1], M.S. Allie[1], M.S. Herbert[1], M.R. Nchodu[1], D.T.L. Jones[2], F.D. Smit[2], R. Nolte[3] and V. Dangendorf[3]
[1]Department of Physics, University of Cape Town, Rondebosch, Cape, 7700, South Africa
[2]iThemba LABS (Laboratory for Accelerator Based Sciences), P.O. Box 722, Somerset West, 7129, South Africa
[3]Physikalisch-Technische Bundesanstalt (PTB), Bundesallee 100, 38116, Braunschweig, Germany

A compact liquid organic neutron spectrometer (CLONS) based on a single NE213 liquid scintillator (5 cm diam. x 5 cm) is described. The spectrometer is designed to measure neutron fluence spectra over the energy range 2-200 MeV and is suitable for use in neutron fields having any type of time structure. Neutron fluence spectra are obtained from measurements of two-parameter distributions (counts versus pulse height and pulse shape) using the Bayesian unfolding code MAXED. Calibration and test measurements made using a pulsed neutron beam with a continuous energy spectrum are described and the application of the spectrometer to radiation dose measurements is discussed.

## INTRODUCTION

Techniques for neutron spectrometry by means of the unfolding of scintillation pulse height spectra measurements have advanced considerably in recent years[1]. Liquid organic scintillators such as NE213, BC501A and EJ301 are popular detectors for this application because of their good pulse height response and their capability to discriminate against γ-rays by means of pulse shape[2]. However, organic scintillators are less convenient for the pulse height spectrum unfolding method at high neutron energies (>100 MeV) because detectors of linear dimensions >15 cm are required to ensure that the escape of high-energy charged particles (especially forward recoil protons) from the detector is not excessive. A large fraction of the high-energy particles need to be brought to rest in the scintillator in order for the pulse height response matrix of the detector to be sensitive to the incident neutron energy and thus be suitable for use in pulse height spectrum unfolding.

In this paper we describe an investigation of an alternative unfolding approach based on the use of both pulse height ($L$) and pulse shape ($S$) measurements from a liquid organic scintillator. The $L$- and $S$-data are used to derive an $LS$-signature in which components due to escaping particles and to different types of charged particles (p, d, α, etc.) that stop in the scintillator are separated. The $LS$-signature is then unfolded, instead of the pulse height spectrum, to determine the neutron spectrum. Escape probabilities for the heavier charged particles, particularly the α-particles, are smaller than for protons. This factor, together with others such as the energy-dependence of the relative intensities of the different components, leads to an $LS$-response matrix which retains its neutron energy-dependence at high energies even when scintillators of small linear dimensions (~5 cm) are used. The need for large detector dimensions in order to perform unfolding at high neutron energies is thus relaxed.

## THE COMPACT LIQUID ORGANIC NEUTRON SPECTROMETER (CLONS)

The prototype CLONS detector consists of a glass-encapsulated NE213 liquid scintillator (5 cm diam. x 5 cm) mounted on a RCA 8575 photomultiplier tube. It was tested in association with experiments carried out by the DOSMAX project at iThemba LABS in November 2003. The neutron beams and monitors used in the DOSMAX experiments were calibrated by the PTB members of the DOSMAX team using a proton recoil telescope and $^{238}$U and $^{235}$U fission ionization chambers.

CLONS was tested using well-collimated neutron beams produced by bombarding natural metallic targets of either lithium (thickness either 5 mm or 8 mm) or beryllium (thickness 13 mm) with a pulsed beam of 200 MeV protons from the iThemba LABS separated-sector cyclotron. The CLONS scintillator was mounted in the neutron beam at a distance of 9.34 m (centre-to-centre) from the Li or Be target. Pulse shape discrimination was accomplished using a Link Systems Model 5010 pulse shape discriminator unit. The outputs from the Link unit were recorded in list mode and then processed off-line[3] to produce a 3-parameter list file of the parameters $L$, $S$ and $T$ (neutron time-of-flight) for events recorded in the run. Four experimental runs were made, one run for each of the Li targets and two for the Be target. The total beam time used for the four runs was 2 hours. The list files obtained in the runs were analysed individually and also in different combinations, one such combination

*Corresponding author: fbrooks@science.uct.ac.za

being an "all-runs-list-file" (ARLF) consisting of the sum of events from all four runs.

Figure 1 shows a plot of counts versus $L$ and $S$ obtained from data selected from the ARLF by a time-of-flight window corresponding to the neutron energy range 180-200 MeV. The structure in the $S$-parameter of this plot is due to the variety of charged particles produced by neutrons interacting with hydrogen and carbon nuclei in the NE213 scintillator and the fact that the scintillation pulse shape depends on the type and energy of the charged particle detected and whether or not this particle is brought to rest in the scintillator. The prominent ridges in the plot are identified as follows: charged particles (mainly protons) that escape from the scintillator (e); and protons (p), deuterons (d) and alpha-particles (α) that come to rest in the scintillator. The continuum (x) between the d- and α-ridges is attributed mainly to multiple neutron scattering and to spallation reactions in which two or more charged particles (p, d, t or τ) are detected simultaneously and thus produce a variety of "non-standard" scintillation pulse shapes.

minima (valleys) between the e- and p-ridges, the p- and d-ridges, and the x-continuum and the α-ridge respectively. The four regions are thus: e ($L$-axis to B1); p (B1 to B2); d+x (B2 to B3) and α (beyond B3). After selecting a threshold value for $L$ the counts in each region were separately projected onto the $L$-axis to produce a pulse height spectrum of 64 channels for the region. The pulse height spectra for the four regions were then assembled into a single array $P$ of 256 channels in which successive 64-channel sectors correspond to the e, p, d+x and α-components respectively. This $P$-array is the $P$-response function of the detector for neutrons of the selected mean energy, for example 190 MeV in the case of Figure 1. The response function is normalised so as to make the sum of counts from $P$ = 1-256 equal to unity.

The same procedure is followed, but without using a $T$-window and without normalisation, to derive the $LS$-signature of an unknown neutron spectrum from the two-parameter measurements of $L$ and $S$.

Normalised $P$-response functions were determined as outlined above (from the ARLF and using the same boundary lines B1-B3) for 46 different neutron energies ($T$-bins) between 8.5 and 190 MeV. The $T$-bins are numbered $n$ = 1-46, from low to high energy, and their widths in terms of energy units vary with the neutron energy. Figure 2 shows a plot of the "$P$-matrix", $R(n,P)$ formed by assembling the 46 normalised $P$-response functions for the different neutron energies. Some important features of the $P$-matrix are the following. The proportions of events in the four components (e, p, d+x and α) vary systematically with neutron energy. At low

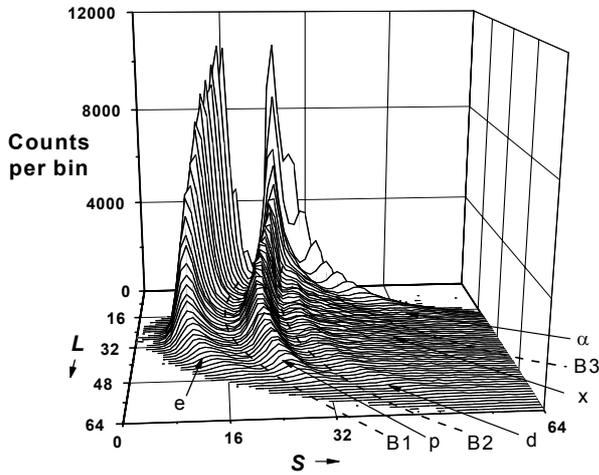

Figure 1: Counts versus pulse height $L$ and pulse shape $S$ for incident neutrons of energy 180-200 MeV. Ridges corresponding to escaping ions (e) and non-escaping protons (p), deuterons (d) and α-particles are indicated. B1-B3 are boundary lines used in the derivation of the $LS$-response functions.

Compton electrons from the detection of γ-rays fall on the e-ridge in Figure 1 but have been suppressed by applying an $LS$-cut that excludes the e-ridge completely for all $L$ less than that corresponding to an 8 MeV electron. The number of Compton electrons that survive this cut is negligible in comparison with the escape component because the intensity of γ-ray background spectra above 8 MeV is very small.

Three boundary lines B1, B2 and B3 (see Figure 1) were used to divide the $LS$-plane into four regions for the derivation of the $LS$-signature. These lines follow the neutron energies the proton component is dominant and the e-component is weak. At the upper energy limit ($n$ = 46) escapes are the largest component. At low-to-medium energies ($n$ = 1-35) structure that varies systematically with neutron energy is clearly evident in the p and the d+x components. The upper limit of $P$ for the α-component (determined by the threshold $R(n,P) > 0.0002$) increases steadily with neutron energy from $n = 1-46$. These features enhance the energy-sensitivity of the ***P***-matrix and are thus important for the unfolding of *LS*-signatures.

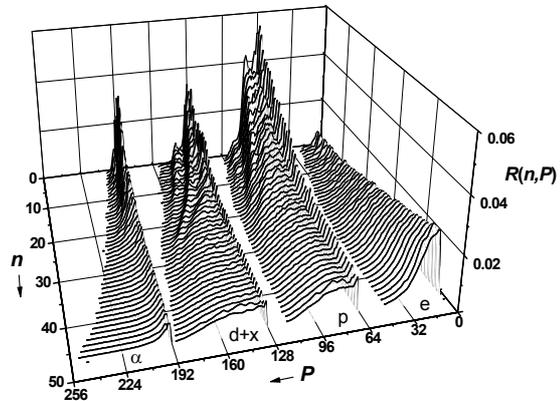

Figure 2: $P$-matrix $R(n, P)$ of the CLONS detector for incident neutrons of energy 8.5-190 MeV ($n$ = 1-46). The response functions for each neutron energy bin ($n$) are normalised so as to make $\Sigma R(n, P) = 1$ for the sum taken over $P$ = 1-256. A threshold level of $R(n, P) > 0.0002$ was used in making the plot.

DETERMINATION OF NEUTRON SPECTRA

Apart from two differences, the unfolding procedure used to obtain neutron spectra from measurements made using the CLONS is the same as that used to obtain neutron spectra by unfolding pulse height spectra. The differences are: firstly, an *LS*-signature is unfolded instead of a pulse height spectrum; and secondly the pulse height response matrix is replaced by the ***P***-matrix. The Bayesian unfolding code MAXED[4] was used to unfold *LS*-signatures derived from experimental data measured using CLONS and the resulting unfolded neutron spectra were compared with neutron spectra calculated independently from the time-of-flight data obtained simultaneously with the *L*- and *S*-measurements. Flat distributions (see Figure 3a) were used as default (*a-priori*) spectra for the MAXED unfolding.

Figure 3a shows the spectral distribution $C(E)$ of neutron counts (per unit energy and unit beam area) obtained from one of the measurements made using the 13 mm thick Be target. The corresponding spectral fluence $\Phi_E(E)$ is shown in Figure 3b. The spectral fluence $\Phi_E(E)$ is given by $\Phi_E(E) = C(E)/\varepsilon(E)$ where $\varepsilon(E)$ is the neutron detection efficiency and $C(E)$ is given by $C(E) = N(n)/A\Delta E(n)$. The index $n$ denotes the energy bin of mean energy $E$, $N(n)$ is the number of counts in bin $n$, determined from either the time-of-flight measurement or by unfolding, $A$ is the cross-sectional area of the detector that is covered by the neutron beam and $\Delta E(n)$ is the energy width of bin $n$. The neutron detection efficiency $\varepsilon(E)$ was determined from $\varepsilon(E) = C(E)/\Phi_E(E)$ using the time-of-flight measurement of $C(E)$ for the Be target (Figure 3a) and the spectral fluence $\Phi_E(E)$ determined for the same target and beam conditions in the DOSMAX calibration runs. All runs were normalised to the same number of monitor counts, based on the neutron beam monitors used in the DOSMAX run. Figure 3c shows the spectral fluences similarly determined for the 8 mm thick Li target.

Figure 4 shows results from MAXED unfolding of three further test spectra obtained by applying various arbitrary time-of-flight profiles to the data of the ARLF. The three profiles defined: (a) a "delta function" corresponding to a mean neutron energy of 105 MeV; (b) a broad distribution of the same mean neutron energy; and (c) a two-peak spectrum with a continuous energy distribution between the peaks. The procedure for these tests was as follows: (i) a subset of events conforming to the specified profile in the *T*-parameter was selected from the ARLF; (ii) an *LS*-signature was formed from the *L* and *S* parameters for this subset; and (iii) this signature was unfolded using MAXED. Figures 4a-c show spectral fluences corresponding to the selected time-of-flight profiles (histograms) and the results obtained from the MAXED unfolding (points) for the three test runs.

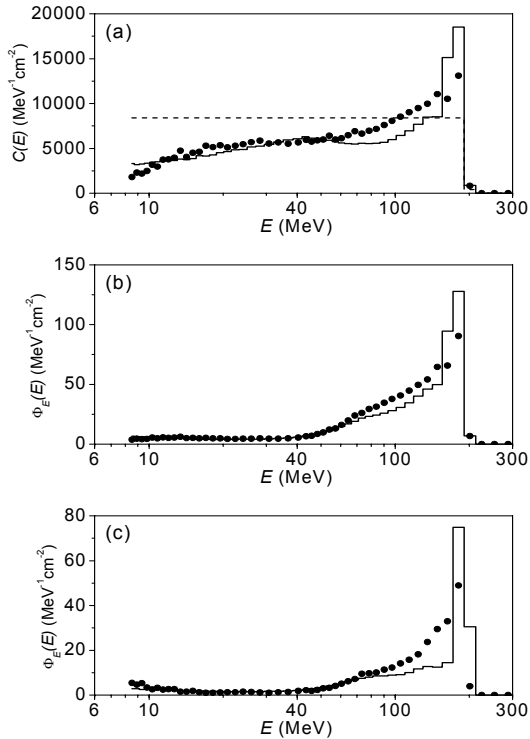

Figure 3: (a) Spectral distributions $C(E)$ of neutron counts (per unit energy and unit beam area) for the 13 mm thick Be target, determined by time-of-flight (histogram) and by unfolding (points). The dashed line shows the default (*a-priori*) spectrum used in the MAXED unfolding. (b) Spectral fluence $\Phi_E(E)$ from the 13 mm thick Be target, determined by time-of-flight (histogram) and by unfolding (points). (c) Spectral fluence $\Phi_E(E)$ determined as in (b), but from the 8 mm thick Li target.

DISCUSSION AND CONCLUSIONS

For the experimental measurements made using the Be and Li targets (Figures 3b and 3c) the spectra obtained by unfolding agree reasonably well with the corresponding time-of-flight spectra at low energies, $E < 60$ MeV, but not so well at higher energies. For the test spectra shown in Figure 4 the agreement between the unfolded spectra and corresponding time-of-flight spectra is also better at lower neutron energy than at high energy. The relatively poor performance of CLONS in determining spectra at the higher energies can probably be attributed largely to insufficient statistical accuracy in the response matrix (see Figure 2) and, to a lesser extent, to the limited statistical accuracy of the *LS*-signatures used in the unfolding. Since the total beam time used in this work was only two hours both of these problems could probably be alleviated by making longer runs. The choice of neutron production targets to be used in future runs could also be improved by including an additional target such as carbon (10-20 mm thick), to provide a more equi-energy neutron spectrum in the ARLF.

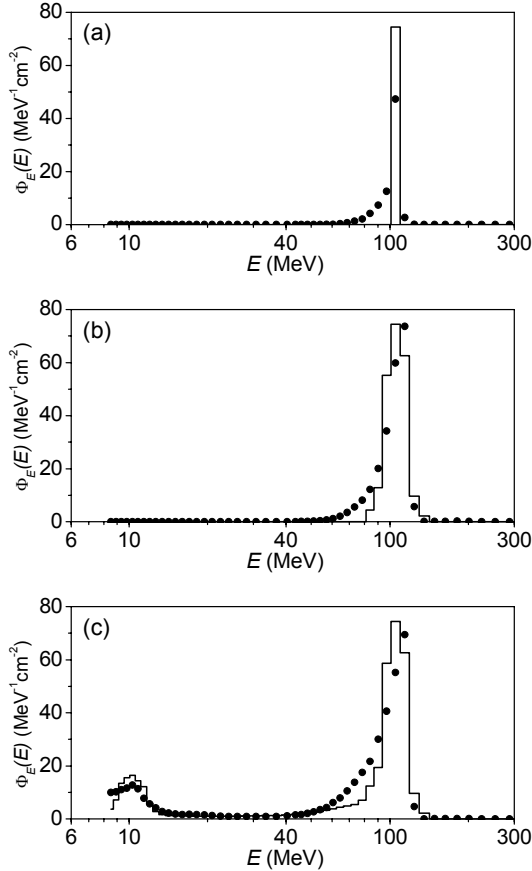

Figure 4: Spectral fluences $\Phi_E(E)$ (points) determined from MAXED unfolding of three arbitrary test spectra (see text) corresponding to: (a) a "delta function" at $E = 105$ MeV; (b) a broad, Gaussian-like peak; and (c) a double-peak spectrum with continuum between the peaks. The histograms show spectral fluences corresponding to the time-of-flight profiles used to select the test spectra from the ARLF.

Ambient dose equivalents $H^*(10)$ corresponding to the spectral fluences shown in figures 3 and 4 have been calculated from $H^*(10) = \Sigma\{h_\Phi(10)\Phi_E(E)\Delta E(n)\}$ where $h_\Phi(10)$ is the fluence-to-dose-equivalent conversion coefficient[5] and the sum is taken over the measured energy range ($n$ = 1-46). Dose equivalents $H_U$ calculated from the spectral fluences determined from unfolding were compared with values $H_T$ determined from the corresponding time-of-flight measurements by calculating fractional deviations $\varphi_H = (H_U - H_T)/H_T$. The values of $\varphi_H$ calculated from the spectra shown in Figures 3b, 3c and 4a-c are 0.046, 0.049, 0.092, 0.054 and 0.036 respectively, thus showing that he CLONS prototype is capable of determining ambient dose equivalents to an accuracy of 4-9% for neutron spectra in the range 9-190 MeV.

The test measurements described in this investigation were limited to the neutron energy range 9-190 MeV by the dynamic range attainable using the Link 5010 pulse shape discriminator, an analogue electronic instrument. The spectrum unfolding method is now widely used for neutron spectrometry in the neutron energy range 2-15 MeV[1]. By using different electronics, including flash ADCs and on-line digital processing, for example, it should be possible to improve the dynamic range of the CLONS so as to cover energies from 1 to 200 MeV in a single compact and simple instrument. Such an instrument would require only a single initial calibration run at a pulsed neutron facility such as iThemba LABS. After calibration the spectrometer will be suitable for operation in neutron fields having any type of time structure. The right cylinder geometry of the NE213 scintillator of the CLONS ensures that its response is approximately independent of the incident neutron direction. A spherical scintillator can be used if a more isotropic response is required.

ACKNOWLEDGEMENTS

We thank Drs H. Klein and H. Schuhmacher for their support in this work, Dr M. Reginatto for providing the MAXED code and the participants in the DOSMAX experiments of November 2003, together with the operating staff of the iThemba LABS cyclotron for their cooperation in this project.